\documentclass[preprint,showpacs,preprintnumbers,amsmath,amssymb]{revtex4}

\usepackage{graphicx}

\begin{document}

\title{
Gravitational bending angle of light for finite distance 
and the Gauss-Bonnet theorem
}
\author{Asahi Ishihara}
\author{Yusuke Suzuki}
\author{Toshiaki Ono}
\author{Takao Kitamura}
\author{Hideki Asada} 
\affiliation{
Faculty of Science and Technology, Hirosaki University,
Hirosaki 036-8561, Japan} 
\date{\today}

\begin{abstract} 
We discuss a possible extension of calculations of 
the bending angle of light in a static, spherically symmetric 
and asymptotically flat spacetime to a non-asymptotically flat case. 
We examine a relation between the bending angle of light 
and the Gauss-Bonnet theorem by using the optical metric. 
A correspondence between the deflection angle of light and 
the surface integral of the Gaussian curvature 
may allow us to take account of 
the finite distance from a lens object 
to a light source and a receiver. 
Using this relation, we propose a method for calculating 
the bending angle of light for such cases. 
Finally, this method is applied to two examples 
of the non-asymptotically flat spacetimes 
to suggest finite-distance corrections: 
Kottler (Schwarzschild-de Sitter) solution to the Einstein equation 
and an exact solution in Weyl conformal gravity. 
\end{abstract}

\pacs{04.40.-b, 95.30.Sf, 98.62.Sb}

\maketitle

\section{Introduction}
The gravitational bending of light by mass 
led to the first experimental confirmations 
of the theory of general relativity. 
In modern astronomy and cosmology, 
the gravitational lensing is widely used as one of 
the important tools for probing extrasolar planets, 
dark matter and dark energy. 

The light bending is also of theoretical importance, 
especially for studying a null structure of a spacetime. 
A rigorous form of the bending angle plays an important role 
in understanding properly a strong gravitational field  
\cite{Frittelli, VE2000, Virbhadra, VNC, VE2002, VK2008, ERT, Perlick}. 
For example, 
strong gravitational lensing in a Schwarzschild black hole 
was considered by Frittelli, Kling and Newman \cite{Frittelli},  
by Virbhadra and Ellis \cite{VE2000} 
and more comprehensively by Virbhadra \cite{Virbhadra}; 
Virbhadra, Narasimha and Chitre \cite{VNC}
studied distinctive lensing features of naked singularities. 
Virbhadra and Ellis \cite{VE2002} 
and Virbhadra and Keeton \cite{VK2008} 
later described 
the strong gravitational lensing by naked singularities; 
DeAndrea and Alexander \cite{DA} discussed the lensing 
by naked singularities to test the cosmic censorship  hypothesis;
Eiroa, Romero and Torres \cite{ERT} treated 
Reissner-Nordstr\"om black hole lensing; 
Perlick \cite{Perlick} discussed the lensing 
by a Barriola-Vilenkin monopole 
and also that by an Ellis wormhole. 
Kitamura, Nakajima and Asada proposed a lens model 
whose gravitational potential declines as $1/r^n$ \cite{Kitamura} 
in order to study the gravitational lensing by exotic matter (or energy) 
\cite{Tsukamoto,Izumi,Kitamura2014,Nakajima2014}
that might follow a non-standard equation of state. 
See Tsukamoto et. al. (2015) \cite{Tsukamoto2014} 
for its possible connection to the Tangherlini solution 
to the higher-dimensional Einstein equation. 

Some recent papers give the expressions for the deflection of light 
for the Kottler (often called Schwarzschild-de Sitter) 
spacetime \cite{Kottler,Lake,RI,Park,Sereno,Bhadra,Simpson,AK} 
and 
for the spherical, static and vacuum exact solution 
in Weyl conformal gravity \cite{MK,Riegert,Edery,Sultana,Cattani}. 
However, their results are not in agreement with each other 
and hence they are controversial. 
The apparent inconsistency among the previous works 
might be caused, because the spacetimes are not asymptotically flat 
and their methods are no longer appropriate 
for treating such a non-asymptotically flat spacetime. 
In the non-asymptotically flat spacetime, 
we can never assume that the source of light 
is located at infinite distance from a gravitational lens object. 
The main purpose of this paper is 
to discuss a possible extension of calculations of 
the bending angle of light in a static, spherically symmetric 
and asymptotically flat spacetime, 
particularly in order to find finite-distance corrections.

For this purpose, we shall examine a relation 
between the bending angle of light 
and the Gauss-Bonnet theorem in differential geometry. 
In this sense, the present paper may 
discuss a possible  
extension of Gibbons and Werner (2008) 
\cite{GW2008}. 
They considered two different domains: one
(say, $\cal{D}$) bounded by two light rays, 
to exhibit the connection between topology and multiple images; 
and the other (say, $\cal{D}^{\prime}$) bounded 
by one light ray and a non-geodesic circular arc, 
to compute the asymptotic deflection angle. 
They suggested that the asymptotic deflection angle of light 
can be written as the surface integral of the Gaussian curvature 
over the domain $\cal D^{\prime}$. 
They did integrate only for the asymptotic case, 
for which they assumed the observer and source are 
in the asymptotically Euclidean region. 
Namely, the angles at the location of the observer and source 
are defined only in Euclidean space \cite{GW2008}.

Throughout this paper, we use the unit of $G=c=1$. 
In the following, the observer may be called the receiver 
in order to avoid a confusion between $r_O$ and $r_0$ by using $r_R$.

\section{Light Propagation, optical metric and Gauss-Bonnet theorem}

\subsection{Static and spherically symmetric spacetime}
We consider a static and spherically symmetric (SSS) spacetime. 
The SSS spacetime can be described as 
\begin{eqnarray}
ds^2 &=& g_{\mu\nu} dx^{\mu} dx^{\nu} 
\nonumber\\
&=& g_{tt}(r) dt^2 + g_{rr}(r) dr^2 + r^2 d\Omega^2 , 
\label{ds2-SSS}
\end{eqnarray}
where the origin of the spatial coordinates is chosen 
as the location of a lens object, 
$\mu$ and $\nu$ run from $0$ to $3$, 
and $d\Omega^2 \equiv d\theta^2 + \sin^2\theta d\phi^2$. 
By introducing two functions as $A(r) \equiv -g_{tt}$ and 
$B(r) \equiv g_{rr}$, 
Eq. (\ref{ds2-SSS}) is rewritten as 
\begin{eqnarray}
ds^2 &=& -A(r) dt^2 + B(r) dr^2 + r^2 d\Omega^2 . 
\label{ds2-SSS-AB}
\end{eqnarray}

\subsection{Optical metric}
Light rays satisfy the null condition as $ds^2 = 0$, 
which is rearranged as, via Eq. (\ref{ds2-SSS-AB}), 
\begin{eqnarray}
dt^2 &=& \gamma_{ij} dx^i dx^j 
\nonumber\\
&=& \frac{B(r)}{A(r)} dr^2 + \frac{r^2}{A(r)} d\Omega^2 ,  
\label{gamma}
\end{eqnarray}
where $i$ and $j$ denote $1$, $2$ and $3$, and 
$\gamma_{ij}$ is often called the optical metric \cite{gamma}. 
The optical metric defines a three-dimensional Riemannian space 
(denoted as $M^{\mbox{opt}}$), 
in which the light ray is expressed as a spatial curve. 

For the spherically symmetric spacetime, 
without the loss of generality, we can choose 
the photon orbital plane as the equatorial plane 
($\theta = \pi/2$). 
The two-dimensional coordinates on the equatorial plane 
are denoted as $x^I$ ($I = 1, 2$), 
where $I$ may mean $r$, $\phi$ particularly in the polar coordinates. 
The nonvanishing components of the optical metric are  
\begin{eqnarray}
\gamma_{rr} &=& \frac{B(r)}{A(r)} , 
\label{gamma-1}
\\
\gamma_{\phi\phi} &=& \frac{r^2}{A(r)} . 
\label{gamma-2}
\end{eqnarray}

Let us suppose the tangent vector field along the light ray. 
The unit tangential vector along the light ray in $M^{\mbox{opt}}$ 
can be defined as 
\begin{equation}
K^I \equiv \frac{dx^I}{dt} . 
\label{K}
\end{equation}
This is a spatial vector. 
Note that $K^I$ is defined in terms of $\gamma_{IJ}$ 
but not $g_{IJ}$, because we consider light rays. 

\subsection{Impact parameter}
In the SSS spacetime, there are two constants of motion 
for a massless particle such as a photon. 
They are the specific energy and the specific angular momentum as 
\begin{eqnarray}
E &=& A(r) \frac{dt}{d\lambda} , 
\\
L &=& r^2 \frac{d\phi}{d\lambda} , 
\end{eqnarray}
where $\lambda$ denotes the affine parameter along the light ray. 
As usual, we define the impact parameter of the light ray as 
\begin{eqnarray}
b &\equiv& \frac{L}{E} 
\nonumber\\
&=& \frac{r^2}{A(r)} \frac{d\phi}{dt} . 
\label{b}
\end{eqnarray}

In terms of the impact parameter and the metric components 
at the position of the massless particle, 
the components of $K^I$ can be expressed as 
\begin{eqnarray}
(K^r, K^{\phi}) = \frac{b A(r)}{r^2} 
\left( \frac{dr}{d\phi}, 1 \right) . 
\end{eqnarray}
Here, the unity of the vector $K^I$ leads to the orbit equation 
as 
\begin{eqnarray}
\left( \frac{dr}{d\phi} \right)^2 
+ \frac{r^2}{B(r)} 
= \frac{r^4}{b^2 A(r)B(r)} . 
\label{orbiteq}
\end{eqnarray}
This can be also derived directly from $ds^2 = 0$. 

Is it safe for us to call what is defined as $b$ 
the impact parameter of light? 
Let us briefly mention this. 
If there were no lens objects, 
then the spacetime would be Minkowskian, 
namely $A(r) = 1$ and $B(r) = 1$ in the polar coordinates, 
and 
$b$ would thus equal to the closest distance 
according to Eq. (\ref{orbiteq}). 
Therefore, $b$ can be safely called 
the impact parameter of the orbit.

\subsection{Angles}
We can define the dyad as 
\begin{eqnarray}
e^I_{\mbox{rad}} &=& 
\left( \frac{1}{\sqrt{\gamma_{rr}}}, 0 \right) , 
\\
e^I_{\mbox{ang}} &=& 
\left( 0, \frac{1}{\sqrt{\gamma_{\phi\phi}}} \right) , 
\end{eqnarray}
which correspond to the unit vector along the radial direction 
from the center of the lens object 
and that along the angular direction, respectively.  

Let $\Psi$ denote the angle of the light ray measured from 
the radial direction. 
It can be defined by 
\begin{equation}
\cos \Psi \equiv \gamma_{IJ} e^I_{\mbox{rad}} K^J , 
\label{cosPsi}
\end{equation}
where we used that $e^I_{\mbox{rad}}$ and $K^J$ are unit vectors. 
This expression is rewritten more explicitly as 
\begin{eqnarray}
\cos \Psi &=& \gamma_{rr} e^r_{\mbox{rad}} K^r 
\nonumber\\
&=& \frac{\sqrt{\gamma_{rr}} b A(r)}{r^2} \frac{dr}{d\phi} . 
\label{cosPsi2}
\end{eqnarray}
This leads to 
\begin{equation}
\sin\Psi = \frac{b \sqrt{A(r)}}{r} , 
\label{sinPsi}
\end{equation}
where we used Eq. (\ref{orbiteq}). 

When we want to obtain $\Psi$ at a point in $M^{\mbox{opt}}$, 
$\sin\Psi$ by Eq. (\ref{sinPsi}) is more convenient than 
$\cos\Psi$ by Eq. (\ref{cosPsi2}), 
because $b \sqrt{A(r)}/r$ can be immediately calculated 
but $\cos\Psi$ includes $dr/d\phi$ that requires a more lengthy calculation.

Let $\Psi_R$ and $\Psi_S$ denote the angles that are measured 
at the receiver position and the source position, respectively. 
Moreover, let $\phi_R$ and $\phi_S$ denote 
the longitudes of the receiver and the source, respectively 
\cite{Memo-Scalars}. 
Let $\phi_{RS} \equiv \phi_R - \phi_S$ denote 
the coordinate separation angle
between the receiver and source. From the three angles 
$\Psi_R$, $\Psi_S$ and $\phi_{RS}$, 
let us define 
\begin{equation}
\alpha \equiv \Psi_R - \Psi_S + \phi_{RS} . 
\label{alpha}
\end{equation}
This is a key equation in the present paper. 

Every two points among the three points of the receiver (R), the source (S)
and the lens center (L) are connected by the geodesics 
in the space $M^{\mbox{opt}}$. 
Hence, the three points in a non-Euclidean space 
constitute an embedded triangle (denoted as ${}^R\bigtriangledown^S_L$). 
The above definition of $\alpha$ depends on the three angles. 
Therefore, we might be dissatisfied with the definition of $\alpha$, 
because the comparison of the scalars at spatially distinct points 
such as $R$ and $S$ 
is quite unclear and even questionable. 
Let us examine whether $\alpha$ is well-defined.

\begin{figure}
\includegraphics[width=10cm]{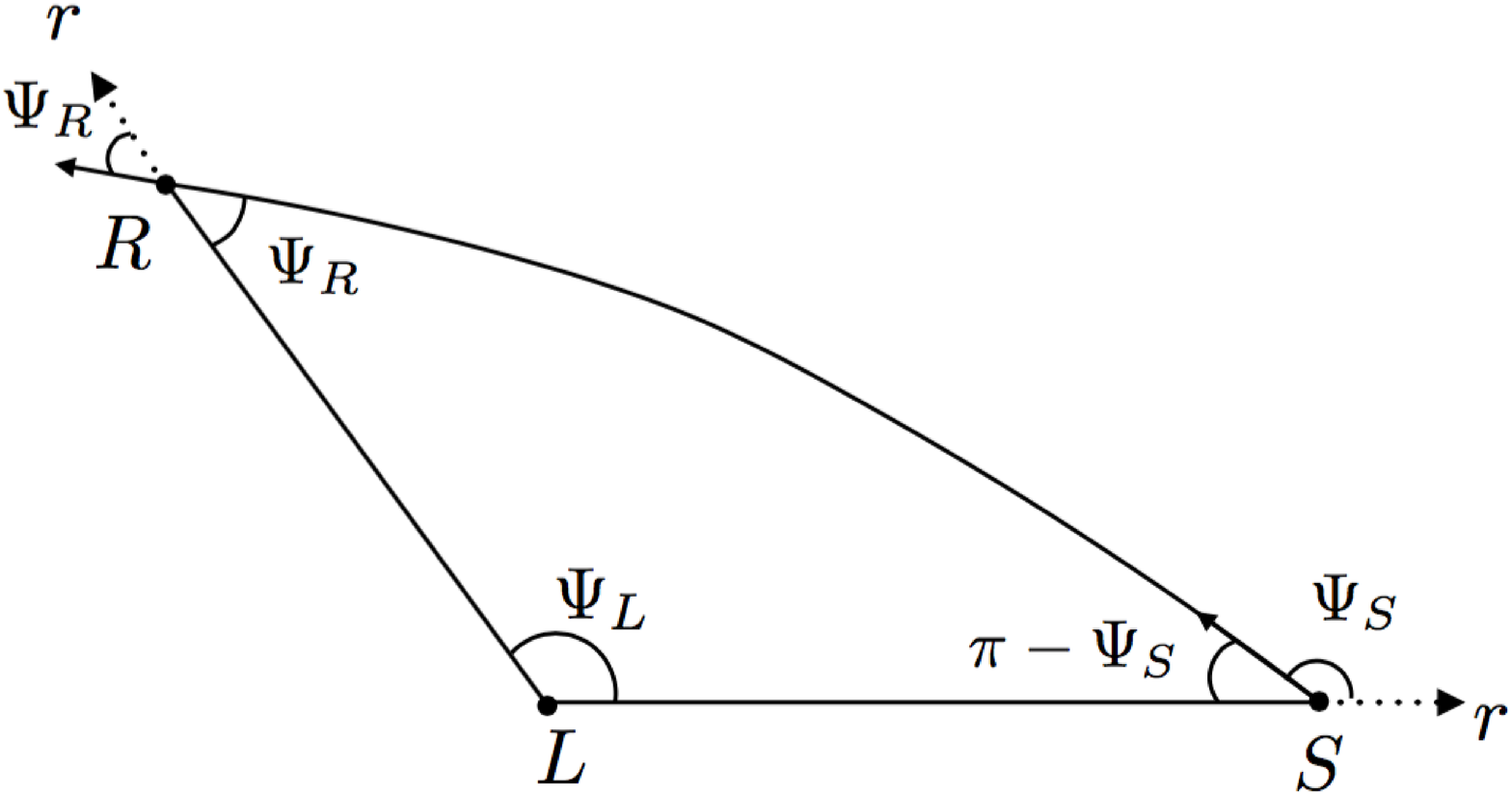}
\includegraphics[width=10cm]{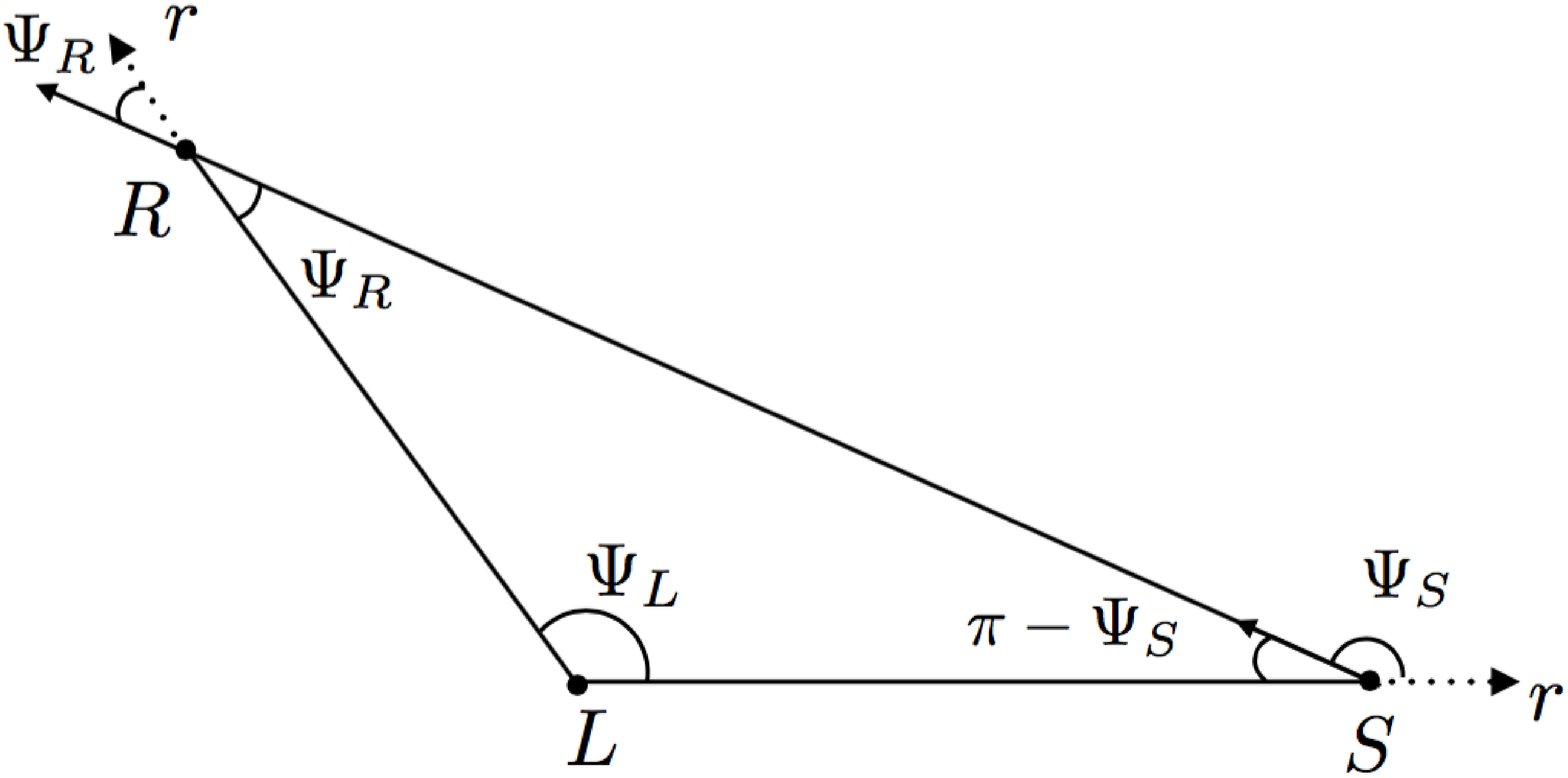}
\caption{
Top: Triangle embedded in a curved space. 
$\alpha_{\Psi}$ does not always vanish. 
Bottom: Triangle in Euclidean space. 
It follows that $\alpha_{\Psi} = 0$. 
}
\label{fig-triangle}
\end{figure}

First, we focus on the triangle ${}^R\bigtriangledown^S_L$. 
Let $\Psi_L$ denote 
the the interior angle at the 
vertex 
$L$. 
The angle $\Psi_S$ is the exterior angle at the 
vertex 
$S$ 
and $\Psi_R$ is the opposite angle of the interior angle at the 
vertex 
$R$ 
by definition.  
Let us define 
\begin{equation}
\alpha_{\Psi} \equiv \Psi_R - \Psi_S + \Psi_L . 
\label{alpha-Psi}
\end{equation}
Note that $\Psi_R$ is the same as the interior angle at $R$. 
See Figure \ref{fig-triangle}. 
Consequently, Eq. (\ref{alpha-Psi}) is rearranged as 
\begin{equation}
\alpha_{\Psi} = \sum_{a=1}^3 \varepsilon_a - \pi , 
\label{alpha-Psi2}
\end{equation}
where $\varepsilon_a$ $(a=1, 2$ and $3$) mean 
the interior angles in the triangle ${}^R\bigtriangledown^S_L$.

If the space $M^{\mbox{opt}}$ is flat, 
it follows that $\alpha_{\Psi} = 0$. 
Hence, this might allow us to interpret $\alpha_{\Psi}$ as a measure of 
the deviation from Euclidean space. 
We shall apply Gauss-Bonnet theorem to the triangle ${}^R\bigtriangledown^S_L$ below. 
 
\subsection{Gauss-Bonnet theorem}
\begin{figure}
\includegraphics[width=12cm]{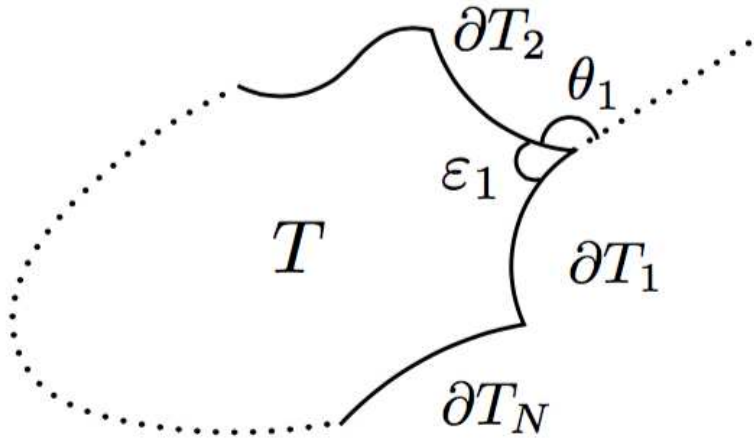}
\caption{
Schematic figure for the Gauss-Bonnet theorem. 
}
\label{fig-GB}
\end{figure}

Suppose that 
$T$ is a two-dimensional orientable surface with boundaries $\partial T_a$ 
($a=1, 2, \cdots, N$) that 
are differentiable curves (See Figure \ref{fig-GB}).  
Let the jump angles between the curves be $\theta_a$ 
($a=1, 2, \cdots, N$). 
Then, the Gauss-Bonnet theorem can be expressed as 
\cite{GB-theorem}
\begin{eqnarray}
\iint_{T} K dS + \sum_{a=1}^N \int_{\partial T_a} \kappa_g d\ell + 
\sum_{a=1}^N \theta_a = 2\pi , 
\label{localGB}
\end{eqnarray}
where 
$K$ denotes the Gaussian curvature of 
the surface $T$, 
$dS$ is the area element of the surface, 
$\kappa_g$ means the geodesic curvature of $\partial T_a$, 
and $\ell$ is the line element along the boundary. 
The sign of the line element is chosen such that it is 
compatible with the orientation of the surface.

By using the Gauss-Bonnet theorem for $N=3$ case, 
Eq. (\ref{alpha-Psi2}) is rewritten as 
\begin{equation}
\alpha_{\Psi} = \iint_{{}^R\bigtriangledown^S_L} K dS 
+ \int_L^S \kappa_g d\ell 
+ \int_S^R \kappa_g d\ell 
+ \int_R^L \kappa_g d\ell , 
\label{alpha-GB1}
\end{equation}
where 
we use 
$\varepsilon_a + \theta_a = \pi$ 
at each point ($a=1, \cdots, N=3$).

For our case, $\kappa_g = 0$ along the boundary curves 
\cite{kappag=0}. 
Therefore, we obtain 
\begin{equation}
\alpha_{\Psi} = \iint_{{}^R\bigtriangledown^S_L} K dS . 
\label{alpha-GB2}
\end{equation}

Eq. (\ref{alpha-GB2}) shows clearly that 
$\alpha_{\Psi}$ is invariant in differential geometry. 
The definition by Eq. (\ref{alpha-Psi}) is thus justified 
\cite{geometrician}. 
The Gaussian curvature can be related with the Riemannian tensor. 
See e.g. Werner (2012) for this relation 
\cite{Werner2012,K-Comment}.

However, it seems impossible to define $\Psi_L$ 
for a case of 
a black hole, because $L$ is the singularity. 
On the other hand, $\phi_{RS}$ seems preferred for practical calculations 
in order to avoid such a problem associated with $\Psi_L$, 
because $\phi_{RS}$ can be defined outside the horizon 
for a black hole case. 

\begin{figure}
\includegraphics[width=12cm]{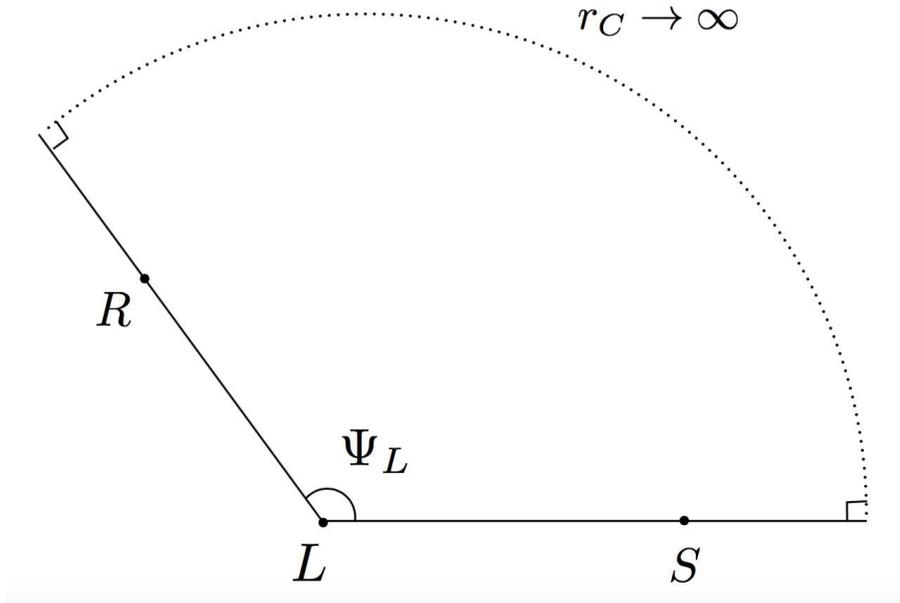}
\caption{
Embedded triangle ${}^{\infty}\bigtriangledown^{\infty}_L$. 
It consists of a circular arc segment $C_r$ 
of coordinate radius $r_C$ centered at the lens 
with taking $r_C \to \infty$, 
and two radial geodesics  
through either the receiver or the source. 
One can determine $\Psi_L$ at the point $L$ from this figure 
by using the Gauss-Bonnet theorem. 
See Eq. (\ref{phiPsi}). 
}
\label{fig-Psi_L}
\end{figure}

We begin by considering another embedded triangle, 
which consists of a circular arc segment $C_r$ 
of coordinate radius $r_C$ centered at the lens 
which intersects the radial geodesic 
through the receiver or the source. 
See Figure \ref{fig-Psi_L}, in which 
we assume the asymptotically flat spacetime 
and a sufficiently large $r_C$, for which 
the embedded triangle is denoted by ${}^{\infty}\bigtriangledown^{\infty}_L$. 
Then, $\kappa_g \to 1/r_C$ and $d\ell \to r_C d\phi$ 
as $r_C \to \infty$ (See e.g. \cite{GW2008}). 
Hence, we obtain 
$\int_{C_r} \kappa_g d\ell \to \phi_{RS}$.  
Applying this result to the Gauss-Bonnet theorem 
for ${}^{\infty}\bigtriangledown^{\infty}_L$ leads to 
\begin{equation}
\Psi_L = \phi_{RS} + \iint_{{}^{\infty}\bigtriangledown^{\infty}_L} K dS . 
\label{phiPsi}
\end{equation}

By substituting Eq. (\ref{phiPsi}) into $\Psi_L$ in $\alpha_{\Psi}$  
of Eq. (\ref{alpha-GB2}), we obtain 
\begin{eqnarray}
\alpha 
&=& \Psi_R - \Psi_S + \phi_{RS} 
\nonumber\\
&=& - \iint_{{}^{\infty}_{R}\Box^{\infty}_{S}} K dS , 
\label{alpha-K}
\end{eqnarray}
where we use Eq. (\ref{alpha}) and 
${}^{\infty}_{R}\Box^{\infty}_{S}$ denotes 
an oriented area 
of ${}^{\infty}\bigtriangledown^{\infty}_L$ subtracted 
by ${}^{R}\bigtriangledown^{S}_L$. 
Eq. (\ref{alpha-K}) shows that $\alpha$ 
is invariant in differential geometry. 
Moreover, it follows that $\alpha=0$ in Euclidean space. 

Both $\alpha$ and $\alpha_{\Psi}$ are geometrically invariant. 
The integration domain for $\alpha_{\Psi}$ includes the lens position. 
Therefore, $\alpha_{\Psi}$ might not be suitable 
for a black hole case. 

On the other hand, it is likely that 
$\alpha$ can avoid such a problem. 
Furthermore, in next section, we shall see that 
(1) $\alpha$ by Eq. (\ref{alpha}) recovers 
the known formula of the bending angle 
for the asymptotic receiver and source 
and 
(2) it can be done in practice to calculate $\alpha$ without encountering 
an infinitely large term for a non-asymptotically flat model, 
though the justification of $\alpha$ by Eq. (\ref{alpha-K}) 
is currently limited within an asymptotically flat case.

\section{Method of calculating the bending angle of light}
There are two ways of calculating $\alpha$, 
because Eq. (\ref{alpha}) always agrees with Eq. (\ref{alpha-K}). 
One method is to use Eq. (\ref{alpha}). 
For this method, all we have to do is to calculate the three angles of 
$\Psi_R$, $\Psi_S$ and $\phi_{RS}$. 
The other method is to use Eq. (\ref{alpha-K}), where 
we first calculate the Gaussian curvature $K$ by using the optical metric 
and next we integrate $K$ over 
the 
quadrilateral 
${}^{\infty}_{R}\Box^{\infty}_{S}$. 
Note that the integration domain ${}^{\infty}_{R}\Box^{\infty}_{S}$, 
especially an expression of the geodesic curve from $S$ to $R$, 
is unknown a priori and hence it must be looked for, 
though the calculation must be straightforward but tedious. 
Let us suppose 
an asymptotic receiver and source of light 
in the Schwarzschild spacetime for instance. 
Even in this case, it is a quite elaborate task to 
calculate the surface integral to recover the known formula 
$\alpha = 4M/b$. 
See Gibbons ans Werner (2008) \cite{GW2008}. 
As a result, it is likely that the first method is much easier 
than the second one.

\subsection{Asymptotically flat case} 
Let us consider the case of the asymptotic flatness. 
Then, we can assume $A(r) \to 1$ and $B(r) \to 1$ as $r \to \infty$. 
As usual, we assume also that the source and receiver 
are located at the null infinity. 
Namely, we assume $r_R \to \infty$ and $r_S \to \infty$. 
Then, let us examine whether Eq. (\ref{alpha}) can recover the textbook 
formula for the deflection angle of light. 
For this purpose, we assume $\Psi_R = 0$ and $\Psi_S = \pi$, 
because we keep $b$ constant with $r_R \to \infty$ and $r_S \to \infty$.

Hence, we obtain 
\begin{equation}
\alpha = \phi_{RS} - \pi . 
\label{alpha-AF}
\end{equation}
All we have to do is to compute $\phi_{RS}$. 

The orbit equation for the light ray 
in the SSS spacetime 
is in a general form as 
\begin{eqnarray}
\left( \frac{du}{d\phi} \right)^2 &=& F(u) , 
\label{F}
\end{eqnarray}
where $u$ denotes the inverse of $r$. 
Please see Eq. (\ref{orbiteq}) for more detail. 

Integrating Eq. (\ref{F}) leads to the angle $\phi_{RS}$ as 
\begin{eqnarray}
\phi_{RS} &=& 
2 \int_0^{u_0} \frac{du}{\sqrt{F(u)}} , 
\label{phiRS}
\end{eqnarray}
where $u_0$ is the inverse of the closest approach 
(often denoted as $r_0$). 
Therefore, substituting this into Eq. (\ref{alpha}) gives 
\begin{equation}
\alpha = 2 \int_0^{u_0} \frac{du}{\sqrt{F(u)}} - \pi . 
\label{alpha-AF}
\end{equation}

\begin{figure}
\includegraphics[width=12cm]{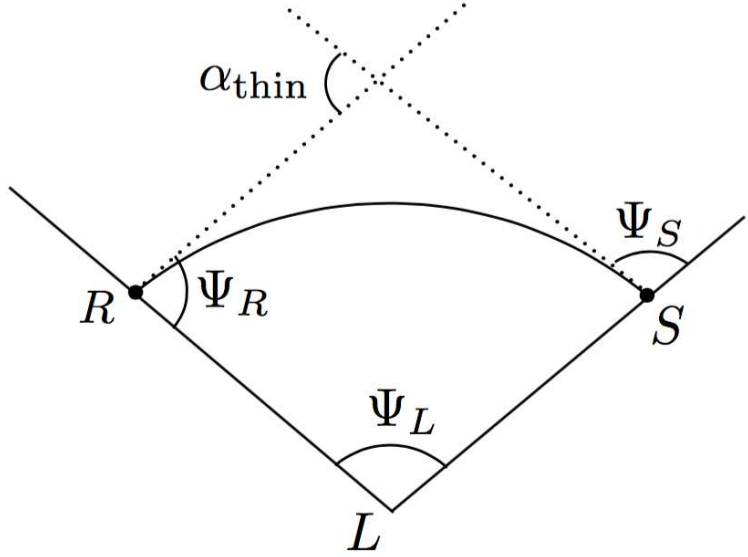}
\caption{ 
Thin lens approximation. 
In this approximation, the light ray deflects only at the lens plane. 
Namely, we assume that the spacetime is flat 
except for the location of the thin lens, 
so that $\Psi_L$ in this figure can be identified with $\phi_{RS}$. 
Let $\alpha_{\mbox{thin}}$ denote the deflection angle of light 
in the thin lens approximation. 
The dotted straight lines are tangential to the light ray 
at the receiver or at the source. 
For the 
quadrilateral 
in Euclidean space, 
$\alpha_{\mbox{thin}} = \Psi_R - \Psi_S + \phi_{RS}$, 
because the sum of the inner angles is $2\pi$. 
}
\label{fig-thin}
\end{figure}

This is exactly the deflection angle of light 
in the literature. 
Therefore, $\alpha$ may be interpreted as 
the deflection angle of light. 
See also Figure \ref{fig-thin} for the thin lens approximation.  
One can see that $\alpha$ is likely to correspond to $\alpha_{\mbox{thin}}$, 
where $\alpha_{\mbox{thin}}$ denotes the deflection angle of light 
in the thin lens approximation.

\subsection{Finite distance cases}
In practice, the thin lens approximation works well 
for most cases in astronomy so far. 
This approximation is almost the same as 
an assumption that a light source and a receiver are 
nearly at the null infinity in the asymptotically flat case. 
To be more specific, 
the present paper assumes that 
the distance from the source to the receiver is finite 
because every observed stars and galaxies are 
located at finite distance from us 
(e.g., at finite redshift in cosmology) 
and the distance is much larger than the size of the lens. 
Hence, we keep $r_R$ and $r_S$ finite. 
Then, let $u_R$ and $u_S$ denote the inverse of 
$r_R$ and $r_S$, respectively. 
Eq. (\ref{alpha}) becomes 
\begin{equation}
\alpha = \int_{u_R}^{u_0} \frac{du}{\sqrt{F(u)}} 
+ \int_{u_S}^{u_0} \frac{du}{\sqrt{F(u)}} 
+\Psi_R - \Psi_S . 
\label{alpha-finite}
\end{equation}
Eq. (\ref{alpha-AF}) is thus corrected.

Eq. (\ref{alpha}), Eq. (\ref{alpha-K}) and Eq. (\ref{alpha-finite}) 
are equivalent to each other. 
They are different from the deflection angle 
that is often used (or argued) in the recent papers \cite{previous}. 

For the Schwarzschild spacetime, the line element becomes 
\begin{eqnarray}
ds^2 &=& -\left( 1 - \frac{r_g}{r} \right) 
dt^2 
+ \frac{dr^2}{\displaystyle 1 - \frac{r_g}{r}} 
\nonumber\\
&&
+ r^2 (d\theta^2 + \sin^2\theta d\phi^2) .  
\label{ds-Kottler}
\end{eqnarray}
Then, $F(u)$ is 
\begin{equation}
F(u) = \frac{1}{b^2} - u^2 + r_g u^3 . 
\label{F-Kottler}
\end{equation}

By using Eq. (\ref{sinPsi}), 
$\Psi_R - \Psi_S$ in the Schwarzschild spacetime is expanded 
in a power series 
in 
$r_g$ as 
\begin{eqnarray}
\Psi_R^{\mbox{Sch}} - \Psi_S^{\mbox{Sch}} 
&\equiv& [\arcsin(bu_R) + \arcsin(bu_S) - \pi] 
\nonumber\\
&& - \frac12 b r_g 
\left( \frac{u_R^2}{\sqrt{1 - b^2u_R^2}}
+ \frac{u_S^2}{\sqrt{1 - b^2u_S^2}} \right) 
+ O(b r_g^2 u_S^3, b r_g^2 u_R^3) . 
\label{Psi-Sch}
\end{eqnarray}
It follows that $\Psi_R - \Psi_S$ for the Schwarzschild case 
approaches $\pi$ 
as $u_S \to 0$ and $u_R \to 0$.

\section{Non-asymptotically flat cases}
Finally, we consider a non-asymptotically flat spacetime 
such as the Kottler solution 
to the Einstein equation 
and an exact solution in the Weyl conformal gravity. 
For such cases, we cannot assume 
the source at the past null infinity ($r_S \to \infty$)
nor the receiver at the future null infinity ($r_R \to \infty$), 
because $A(r)$ diverges or does not exist as $r \to \infty$. 
Hence, we should keep the source and the receiver 
to be at finite distance from the lens object. 
It is Eq. (\ref{alpha-finite}) that we can use for such a case. 
As mentioned already,  
Eq. (\ref{sinPsi}) is more convenient for calculating $\Psi_R$ and $\Psi_S$ 
than Eq. (\ref{cosPsi}), 
since Eq. (\ref{sinPsi}) needs 
a local quantity 
but not any derivative.  
For the two cases, the explicit expressions are as follows. 

\subsection{Kottler case}
For the Kottler spacetime \cite{Kottler}, the line element is 
\begin{eqnarray}
ds^2 &=& -\left( 1 - \frac{r_g}{r} - \frac{\Lambda}{3}r^2 \right) 
dt^2 
+ \frac{dr^2}{\displaystyle 1 - \frac{r_g}{r} - \frac{\Lambda}{3}r^2} 
\nonumber\\
&&
+ r^2 (d\theta^2 + \sin^2\theta d\phi^2) , 
\label{ds-Kottler}
\end{eqnarray}
where $\Lambda$ denotes the cosmological constant. 

By using Eq. (\ref{sinPsi}), 
$\Psi_R - \Psi_S$ is expanded in terms of $r_g$ and $\Lambda$ as 
\begin{align}
\Psi_R-\Psi_S
&=\Psi^{Sch}_R-\Psi^{Sch}_{S} 
- \frac{b\Lambda}{6u_R\sqrt{1-b^2u_R^2}} 
- \frac{b\Lambda}{6u_S\sqrt{1-b^2u_S^2}} 
\notag\\
&+ \frac{bu_R(-1+2b^2u_R^2)}{8(1-b^2u_R^2)^{3/2}}
\left(r_g^2u_R^2+\frac{2r_g\Lambda}{3u_R}+\frac{\Lambda^2}{9u_R^4}\right)
\notag\\
&+\frac{bu_S(-1+2b^2u_S^2)}{8(1-b^2u_S^2)^{3/2}}
\left(r_g^2u_S^2+\frac{2r_g\Lambda}{3u_S}
+\frac{\Lambda^2}{9u_S^4}\right) 
\notag\\
&+ O(r_g^3, r_g^2\Lambda, r_g\Lambda^2, \Lambda^3) , 
\label{Psi-Kottler}
\end{align}
where $\Psi_R^{\mbox{Sch}} - \Psi_S^{\mbox{Sch}}$ 
is a part existing in Schwarzschild spacetime.  
Note that the above expansion of $\Psi_R - \Psi_S$ is divergent 
as $u_S \to 0$ and $u_R \to 0$. This is because the spacetime 
is not asymptotically flat and hence it does not 
allow the limit of $u_S \to 0$ and $u_R \to 0$. 
Hence, 
the power series form 
by Eq. (\ref{Psi-Kottler}) 
must be used within a certain finite radius of convergence. 

For the Kottler case, $F(u)$ becomes 
\begin{equation}
F(u) = \frac{1}{b^2} - u^2 + r_g u^3 + \frac{\Lambda}{3} . 
\label{F-Kottler}
\end{equation}
Hence, we obtain 
\begin{align}
\phi_{RS}
=&\pi-\arcsin(bu_{R})-\arcsin(bu_{S})
\notag\\
&+\frac{r_{g}}{b}\left[\frac{1}{\sqrt{1-b^{2}u_{R}^{2}}} 
\left(1-\frac{1}{2}b^{2}u_{R}^{2}\right) 
+ \frac{1}{\sqrt{1-b^{2}u_{S}^{2}}}\left(1-\frac{1}{2}b^{2}u_{S}^{2}\right)\right]
\notag\\
&+\frac{\Lambda b^{3}}{6}\left[\frac{u_{R}}{\sqrt{1-b^{2}u_{R}^{2}}} 
+ \frac{u_{S}}{\sqrt{1-b^{2}u_{S}^{2}}}\right] 
+ \frac{r_{g}\Lambda b}{12}
\left[\frac{2-3b^{2}u_{R}^{2}}{(1-b^{2}u_{R}^{2})^{\frac{3}{2}}}
+\frac{2-3b^{2}u_{S}^{2}}{(1-b^{2}u_{S}^{2})^{\frac{3}{2}}}\right]
+ O(r_{g}^{2},\Lambda^{2}) .
\label{phi-Kottler}
\end{align}

By using Eqs. (\ref{Psi-Kottler}) and (\ref{phi-Kottler}), 
we obtain the correct deflection angle of light as 
\begin{align}
\alpha
=&
\frac{r_{g}}{b}\left[\sqrt{1-b^{2}u_{R}^{2}}+\sqrt{1-b^{2}u_{S}^{2}}\right]
\notag\\
& -\frac{\Lambda b}{6}
\left[\frac{\sqrt{1-b^{2}u_{R}^{2}}}{u_{R}}+\frac{\sqrt{1-b^{2}u_{S}^{2}}}{u_{S}}\right]
\notag\\
& +\frac{r_{g}\Lambda b}{12}\left[\frac{1}{\sqrt{1-b^{2}u_{R}^{2}}}
+\frac{1}{\sqrt{1-b^{2}u_{S}^{2}}}\right] 
+ O(r_{g}^{2},\Lambda^{2}). 
\label{alpha-Kottler}
\end{align}
Some terms in this expression may apparently diverge 
in the limit as both $b u_R \to 0$ and $b u_S \to 0$. 
Note that this limit has no relevance with astronomical observations 
in the Kottler spacetime. 
Therefore, the apparent divergence does not matter. 

Aghili, Bolen and Bombelli have recently discussed 
numerically effects of a slowly varying Hubble parameter on the 
gravitational lensing \cite{Aghili}. 
It is left as a future work to examine an application of 
the present approach to such a cosmological model 
with a slowly varying Hubble parameter.

\subsection{Weyl conformal gravity case}
Weyl conformal gravity introduces three independent parameters 
(often denoted as $\beta$, $\gamma$ and $k$) 
into the spherical solution, for which 
Birkhoff's theorem was proven in conformal gravity 
\cite{Riegert}. 
The line element with the three parameters is \cite{MK}
\begin{eqnarray}
ds^2&=&-A(r) dt^2+\frac{1}{A(r)} dr^2 
+ r^2(d\theta^2+\sin^2\theta d\phi^2), \nonumber\\
A(r)&=&1 - 3m\gamma - \frac{2m}{r} + \gamma r -k r^2 ,
\label{ds-Weyl}
\end{eqnarray}
where we defined  
$m \equiv \beta (2 - 3 \beta\gamma)/2$.
The term with the coefficient $k$ makes the same contribution 
as the cosmological constant in the Kottler spacetime 
that has been studied above. 
Henceforth, 
we omit the $r^2$ term for 
brevity. 

By using Eq. (\ref{sinPsi}), 
$\Psi_R - \Psi_S$ is expanded in a power series 
in  
$\beta$ and $\gamma$ as 
\begin{align}
\Psi_R-\Psi_S \equiv &\Psi_R^{\mbox{Sch}}-\Psi_S^{\mbox{Sch}} \notag\\
&+\frac{b\gamma}{2}\left(\frac{u_R}{\sqrt{1-b^2u^2_R}}+\frac{u_S}{\sqrt{1-b^2u^2_S}}\right) \notag\\
&-\frac{m\gamma}{2}\left[\frac{bu_R(2-b^2u_R^2)}{(1-b^2u_R^2)^{3/2}}+\frac{bu_S(2-b^2u_S^2)}{(1-b^2u_S^2)^{3/2}}\right] +O(m^2,\gamma^2) . 
\label{Psi-Weyl}
\end{align}
Note that this series expansion of $\Psi_R - \Psi_S$ is divergent 
as $u_S \to 0$ and $u_R \to 0$. 
This is because the non-asymptotic flatness of the spacetime 
does not 
allow 
the limit of $u_S \to 0$ and $u_R \to 0$. 
Hence, we must use Eq. (\ref{Psi-Weyl}) 
within its certain radius of convergence. 

For the conformal gravity case with $k=0$, 
$F(u)$ becomes 
\begin{equation}
F(u) = \frac{1}{b^2} - u^2 + 2m u^3 
+ \Gamma u^2 - \gamma u . 
\label{F-Weyl}
\end{equation}
Then, $\phi_{RS}$ is obtained as 
\begin{align}
\phi_{RS}
=& [\pi-\arcsin(bu_R)-\arcsin(bu_S)] 
\notag\\
&
+ \frac{m}{b} \left( \frac{2-b^2u_R^2}{\sqrt{1-b^2u_R^2}} 
+ \frac{2-b^2u_S^2}{\sqrt{1-b^2u_S^2}} \right)
\notag\\
& - \frac{\gamma}2 \left( \frac{b}{\sqrt{1-b^2u_R^2}} + \frac{b}{\sqrt{1-b^2u_R^2}} \right) 
\notag\\ 
& 
+\frac{m\gamma}{2}\left[\frac{b^3u_R^3}{(1-b^2u_R^2)^{3/2}}+\frac{b^3u_S^3}{(1-b^2u_S^2)^{3/2}}\right] 
+ O(m^2,\gamma^2) .
\label{phi-Weyl}
\end{align}

In total, we obtain $\alpha$ for the Weyl conformal gravity case as 
\begin{align}
\alpha=
& \frac{2m}{b} \left(\sqrt{1-b^2u_R^2}+\sqrt{1-b^2u_S^2}\right) 
\notag\\
& -m\gamma\left(\frac{bu_R}{\sqrt{1-b^2u_R^2}}+\frac{bu_S}{\sqrt{1-b^2u_S^2}}\right) + O(m^2,\gamma^2) . 
\label{alpha-Weyl}
\end{align}
The terms linear in $\gamma$ cancel out in the expression for
the deflection angle of light. 
Hence, this might correct the results in previous papers 
\cite{Edery,Sultana,Cattani} 
that reported non-zero contributions from $\gamma$.

\subsection{Far source and receiver}
Finally, let us consider an asymptotic case 
as $b u_S \ll 1$ and $b u_R \ll 1$, 
which mean that both the source and the receiver are very far from 
the lens object. 
Note that $b u_S \to 0$ and $b u_R \to 0$ might cause 
the divergent terms in the deflection angle. 
Hence, we focus on the dominant part of each term 
in a series expansion without taking the limit. 
Let us write down approximate expressions 
for the deflection of light. 

\noindent
(1) Kottler case:\\
It follows that the expression for $\phi_{RS}$ in the far approximation 
coincides with the seventh and eighth terms of Eq. (5) in \cite{Sereno}, 
the third and fifth terms of Eq. (15) in \cite{Bhadra}, 
and the second term of Eq. (14) in \cite{AK}. 
However, they \cite{Sereno,Bhadra,AK} did not consider $\Psi_R - \Psi_S$. 
Eq. (\ref{alpha-Kottler}) becomes 
\begin{equation}
\alpha \sim \frac{2r_g}{b} 
- \frac16 \Lambda b \left( \frac{1}{u_R} + \frac{1}{u_S} \right) 
+ \frac{1}{6} r_g \Lambda b . 
\label{alpha-far-Kottler}
\end{equation}
This might give a correction to the previous results \cite{Sereno,Bhadra,AK}.  
For instance, Sereno (2009) considered only $\phi_{RS}$. 
See \cite{previous} for a more subtle case associated 
with Rindler and Ishak's approach.

\noindent
(2) Weyl conformal gravity case:\\
For the Weyl conformal gravity, the deflection angle of light 
in the far approximation becomes 
\begin{eqnarray}
\alpha
&\sim& 
\frac{4m}{b}  
+ O(m^2,\gamma^2) . 
\label{alpha-Weyl}
\end{eqnarray}
Note that $m\gamma$ parts from $\Psi_R - \Psi_S$ and from $\psi_{RS}$ 
cancel out. 
See Eqs. (\ref{Psi-Weyl}) and (\ref{phi-Weyl}).

\begin{figure}
\includegraphics[width=10cm]{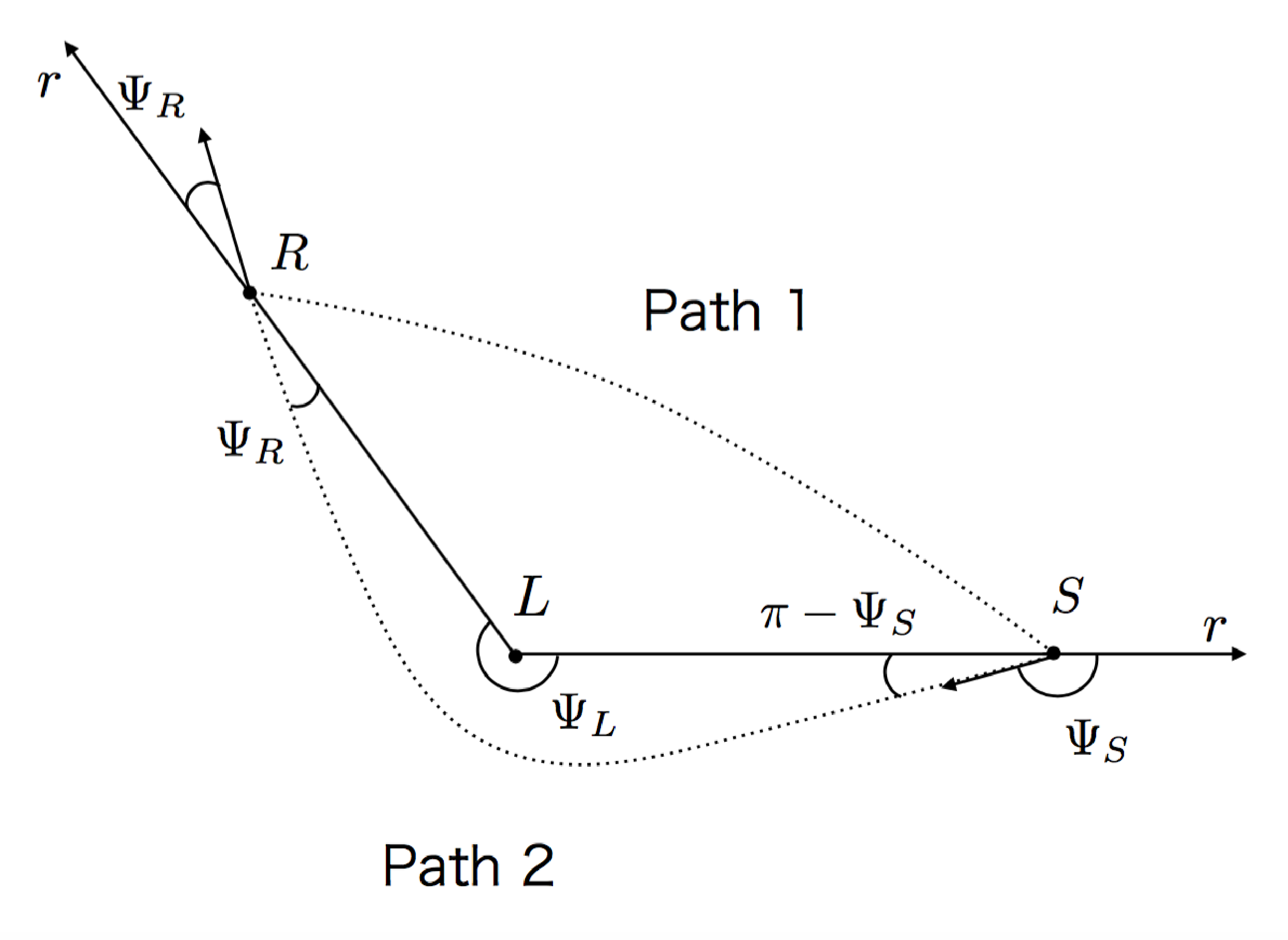}
\caption{ 
Two light paths. 
In the present paper, we focus on the path 1 
in this figure, because it corresponds to the brightest lensed image 
and it plays a crucial role in astronomy. 
There is another possible path (Path 2 in this figure). 
The two light paths are denoted by dotted lines. 
}
\label{fig-path2}
\end{figure}

Before closing this section, 
we briefly mention another light path (Path 2 in Figure \ref{fig-path2}). 
The bending angle for this path is computable, 
if we take account of the orientation of Path 2 (See Fig. \ref{fig-path2}).

\section{Conclusion}
In this paper, we studied a connection between the bending angle of light 
and the Gauss-Bonnet theorem by using the optical metric in the SSS spacetimes. 
A correspondence of the deflection angle of light 
to the surface integral of Gaussian curvature 
may allow us to take account of 
the finite distance of a light source and 
a receiver from a lens object.

The proposed approach of calculating the deflection angle of light 
by Eq. (\ref{alpha}) was applied to two examples 
of the non-asymptotically flat spacetimes: 
Kottler solution to the Einstein equation 
and an exact solution in Weyl conformal gravity. 
For the both cases, 
we suggested finite-distance corrections to the deflection angle of light 
without encountering an infinitely large term, as a conjecture, 
because the justification of $\alpha$ by Eq. (\ref{alpha-K})  
cannot be applied to such a non-asymptotically flat case as it is. 
It would be interesting to examine whether or not the justification 
of Eq. (\ref{alpha}) is extended 
to a non-asymptotically flat case. 
If it is not, we may find new corrections to the deflection angle of light. 
It would be interesting to study along this direction.

Moreover, let us suppose that 
the light ray passes near a relativistic compact object. 
For this case, the deflection angle of light may exceed $2\pi$ 
to procedure the relativistic images. 
For such a large deflection case, 
the orbit has the winding number $W$ 
that may be the unity or more. 
Eqs. (\ref{alpha}) and (\ref{alpha-GB2}) still work, 
because the light ray lives on a single plane in $M^{\mbox{opt}}$ 
\cite{Weyl-Comment}. 
Further study along the direction of the relativistic strong lensing 
by using the present approach is left for future work. 

We are grateful to Marcus Werner for the stimulating discussions, 
especially for his useful comments on the Gauss-Bonnet theorem 
and the earlier version of the manuscript. 
We wish to thank Makoto Sakaki for giving us 
the useful literature information on the Gauss-Bonnet theorem. 
We would like to thank Toshifumi Futamase, Masumi Kasai, 
Yuuiti Sendouda, Ryuichi Takahashi, Koji Izumi, Tomohito Suzuki 
and Takumi Takahashi for the useful conversations. 
This work was supported 
in part by JSPS Grant-in-Aid for Scientific Research, 
(Kiban C) No. 26400262 (H.A.) and 
in part by MEXT (Shingakujutsu) No. 15H00772 (H.A.).

\newpage

\begin{thebibliography}{99}
\bibitem{gamma}
To be more precise, $\gamma_{ij}$ might be 
denoted 
as $\gamma_{ij}^{\mbox{opt}}$, since the notation as $\gamma_{ij}$ 
is often used for the induced metric on a hypersurface 
in the 3+1 formulation in the literature. 
However, we use $\gamma_{ij}$ for denoting the optical metric 
for its simplicity in this paper, 
because the induced metric does not appear in this paper 
and hence readers would not be confused.
\bibitem{Memo-Scalars}
Let us suppose two scalars. 
One scalar $\xi(A)$ is defined 
at 
a point $A$. 
The other scalar $\eta(B)$ is at a very different point $B$. 
The sum of $\xi(A)$ and $\eta(B)$ and the difference between 
$\xi(A)$ and $\eta(B)$ have no invariant meaning in general.
\bibitem{kappag=0}
Fermat's principle in a spacetime is the principle that the light path 
makes the arrival time stationary. 
It is expressed as $\delta\int dt = 0$ along the light ray \cite{Perlick}. 
In our case, hence, Eq. (\ref{gamma}) implies that the orbit of light 
is a geodesic 
in $M^{\mbox{opt}}$ with the optical metric $\gamma_{ij}$.
\bibitem{geometrician}
It is likely that mathematicians prefer to define $\alpha$ 
by Eq. (\ref{alpha-GB2}). 
However, it is quite elaborate in practice to 
perform the areal integral of Eq. (\ref{alpha-GB2}). 
In order to do it, 
we have to find out the integration domain first, 
especially the domain boundary 
connecting the receiver and the source. 
This 
boundary 
is the light ray 
that is unknown a priori for most cases 
and hence the expression of the 
boundary  
can be obtained 
only after the orbit equation is solved.
\bibitem{Weyl-Comment}
It is likely that the nonvanishing contributions to $\alpha$ 
at the linear order of $\Gamma$ and $\gamma$ in the previous works 
are because they did not use the optical metric 
to define the angles at the receiver and the source 
and they employed a different definition of $\alpha$ 
by following \cite{RI}. 
In particular, $\gamma$ does not contribute at the leading order. 
It might be interesting to interpret this thing 
by a projective equivalence \cite{GWW} 
between the solution in the Weyl conformal gravity 
and another spacetime.
\end{thebibliography}
\end{document}